\documentclass[twocolumn,amssymb,aps,superscriptaddress]{revtex4}
\usepackage{graphicx}
\usepackage{dcolumn}
\usepackage{amsmath}
\usepackage{bm}
\usepackage{latexsym,epsfig}
\long\def\symbolfootnote[#1]#2{\begingroup%
\def\thefootnote{\fnsymbol{footnote}}\footnote[#1]{#2}\endgroup}

\newcommand\ir{$6\,^2$D$_{3/2}$\,-\,$7\,^2$P$_{1/2}\,$}
\newcommand\shelf{$6\,^2$D$_{3/2}$\,-\,$7\,^2$P$_{3/2}\,$}
\newcommand\pump{$7\,^2$S$_{1/2}$\,-\,$7\,^2$P$_{1/2}\,$}

\newcommand\dt{$6\,^2$D$_{3/2}\,$}
\newcommand\df{$6\,^2$D$_{5/2}\,$}
\newcommand\pt{$7\,^2$P$_{3/2}\,$}
\newcommand\ph{$7\,^2$P$_{1/2}\,$}
\newcommand\sh{$7\,^2$S$_{1/2}\,$}

\begin{document}

\preprint{\today}

\title{On-line Excited-State Laser Spectroscopy of Trapped Short-Lived Ra$^+$ Ions}
\vspace{0.5cm}
\author{O.\ O.\ Versolato\footnote{\tt versolato@KVI.nl}, G.\ S.\ Giri, L.\ W.\ Wansbeek, J.\ E.\ van\ den\ Berg, D.\ J.\ van\ der\ Hoek,
              K.\ Jungmann, W.\ L.\ Kruithof, C.\ J.\ G.\ Onderwater, B.\ K.\ Sahoo, B.\ Santra, P.\ D.\ Shidling, R.\ G.\ E.\ Timmermans,
              L.\ Willmann, H.\ W.\ Wilschut}
\affiliation{University of Groningen, Kernfysisch Versneller Instituut, NL-9747 AA Groningen, The Netherlands}%
\date{\today}
\vskip1.0cm

\begin{abstract}
As an important step towards an atomic parity violation experiment in one single trapped Ra$^+$ ion, laser spectroscopy experiments were performed with on-line produced short-lived $^{212,213,214}$Ra$^+$ ions. The isotope shift of the \ir
and \shelf transitions and the hyperfine structure constant of the \ph and \dt states in $^{213}$Ra$^+$ were measured.
These values provide a benchmark for the required atomic theory. A lower limit of $232(4)$ ms for the lifetime of the
metastable \df state was measured by optical shelving.
\end{abstract}


\maketitle

The radium ion, Ra$^+$, is a promising candidate for an  atomic parity violation (APV) experiment with one single trapped ion~\cite{fortsonprl93,koerberjpb03,shermanprl05,wansbeekprar08}. APV experiments~\cite{fortsonprl93,koerberjpb03,shermanprl05,wansbeekprar08,Bouchiat1982358,woodsc97,bennettprl99,gwinner2006,stancari2007,PhysRevLett.103.071601} are sensitive probes of the electroweak interaction at low energy. APV is due to the exchange of the $Z^0$ boson between the electrons and the quarks in the atomic nucleus. Its size depends on the mixing angle of the photon and the $Z^0$ boson, which is a fundamental parameter of the electroweak theory. The APV signal is strongly enhanced in heavy atoms~\cite{bouchiatpl74} and it is measurable by exciting suppressed (M1, E2) transitions~\cite{guenampla05}.  The predicted enhancement in Ra$^+$ is about 50 times larger than in Cs atoms~\cite{wansbeekprar08,dzubapra01,PhysRevA.79.062505}, for which the most accurate measurement has been performed~\cite{woodsc97,bennettprl99,PhysRevLett.102.181601}. However, laser spectroscopy on trapped Ra$^+$ ions has not been performed yet, and certain spectroscopic information, needed to test the required atomic many-body theory, is lacking~\cite{wansbeekprar08}. For instance, the lifetimes of the \dt and \df states, which are important quantities for a single-ion APV experiment, have not been measured yet. These states are also relevant for a potential Ra$^+$ optical clock~\cite{dzubapra00,sahoopra07,bijayapra09}.

\begin{figure}[t]
\includegraphics[width = 8.0cm, angle = 0]{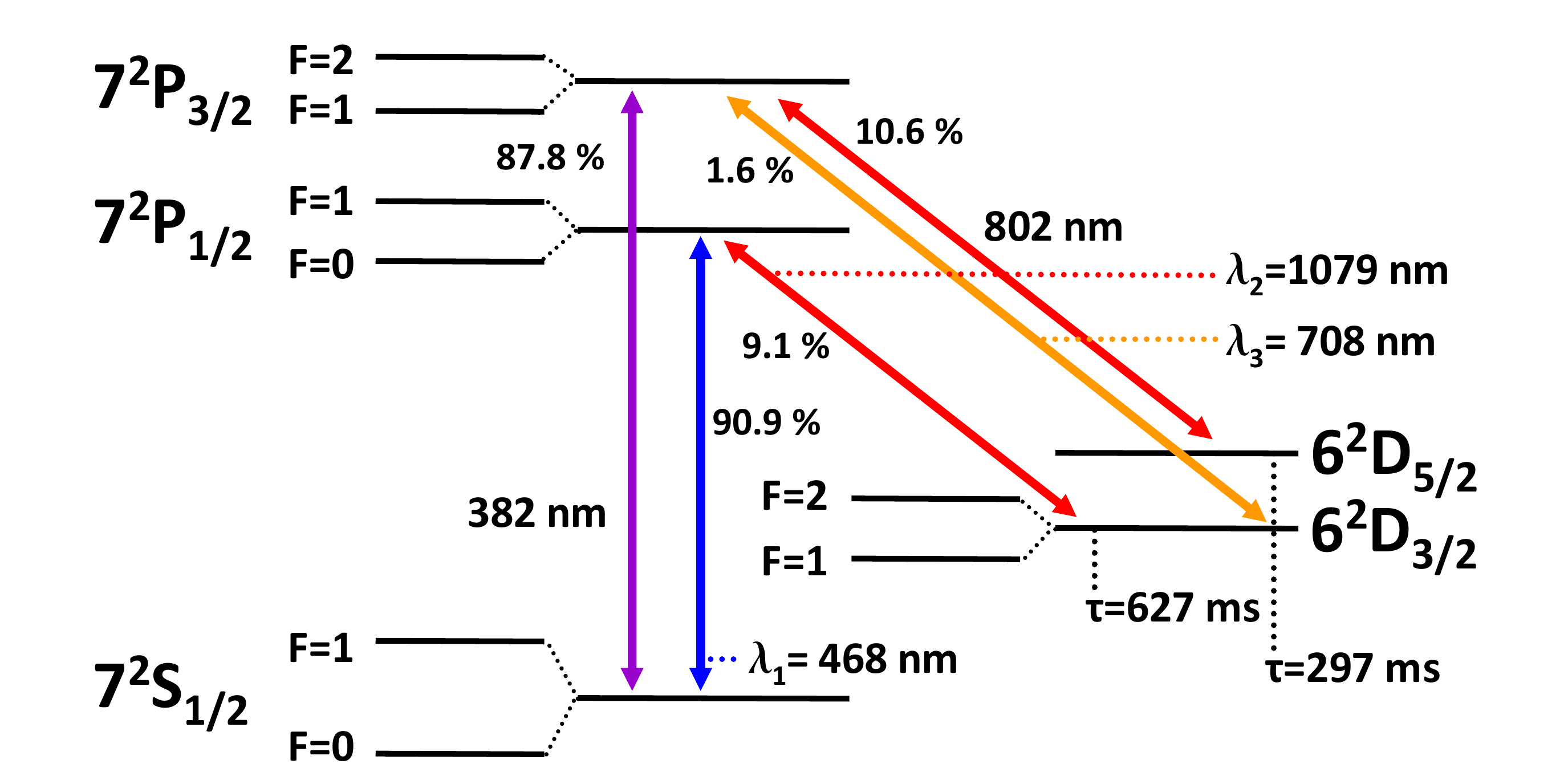}
\caption{$^{213}$Ra$^+$ level scheme with wavelengths from Ref.~\cite{rasmussenzp33} and branching ratios and lifetimes calculated by Ref.~\cite{sahoopra07}.} \label{level}
\end{figure}

Up to now, accurate experimental information on the optical spectrum of Ra$^+$  ({\it cf.} Fig.~ \ref{level}) was only available from measurements at the ISOLDE facility at CERN, where the isotope shift (IS) and hyperfine structure (HFS) of the \sh\!, \ph\!, and \pt\! states were obtained by collinear spectroscopy over a large range of isotopes~\cite{ISOLDE1987isohyper,ISOLDEmomentswPhysRevLett.59.771}. The only absolute measurement of the relevant wavelengths dates back to arc emission spectroscopy performed on $^{226}$Ra$^+$ in 1933~\cite{rasmussenzp33}. We present here the results of on-line excited-state laser spectroscopy experiments of trapped, short-lived $^{212,213,214}$Ra$^+$ ions, obtained at the TRI$\mu$P facility~\cite{Shidling2009305} of the KVI in Groningen. IS and HFS measurements were performed to constrain the atomic theory: HFS is a sensitive probe of the atomic wave functions in the nucleus~\cite{PhysRevA.80.044502}, the accuracy of which is
important for APV, while experiments on different isotopes serve to cancel remaining uncertainties in the atomic theory~\cite{wansbeekprar08}.

\setlength{\floatsep}{12pt}
\setlength{\textfloatsep}{12pt}

\begin{figure}[t]
\includegraphics[width = 7.0cm, angle = 0]{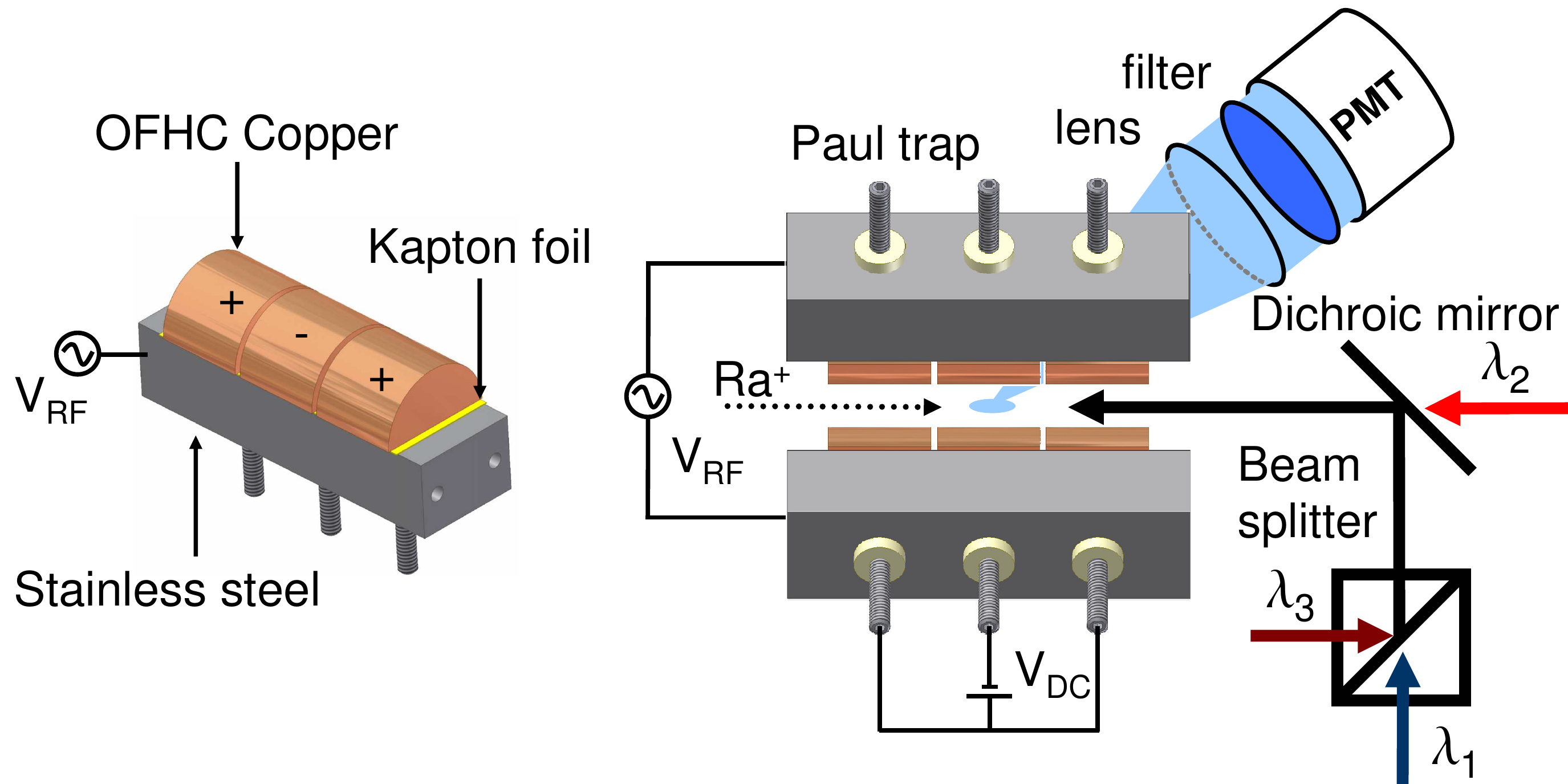}
\caption{Schematic overview of the experimental setup behind the RFQ cooler.} \label{setup}
\end{figure}

Radium isotopes were produced in inverse kinematics by bombarding an 8.5 MeV/nucleon $^{206}$Pb beam of  typically $3 \times 10^{10}$ particles/s from the AGOR cyclotron on a 4 mg/cm$^2$ diamond-like carbon foil, and emerged from the fusion-evaporation reactions $^{206}$Pb + $^{12}$C $\rightarrow$ $^{218-x}$Ra, in which $x$ neutrons were liberated. The isotopes $^{212}$Ra, $^{213}$Ra, and $^{214}$Ra were separated from the primary beam and fission products in the magnetic separator~\cite{Berg2006169}. They were stopped and re-ionized to Ra$^+$ in a Thermal Ionizer (TI)~\cite{Shidling2009305} with a transmission efficiency of 8\%. Rates of 800 $^{212}$Ra$^+$/s, 2600 $^{213}$Ra$^+$/s, and 1000 $^{214}$Ra$^+$/s were extracted as an ion beam with an energy of 2.8 keV. The Ra$^+$ isotopes were passed through a Wien Filter (which eliminated contaminants from the TI), and electrostatically decelerated upon injection in a (N$_2$ or Ne) gas-filled Radio Frequency Quadrupole (RFQ) cooler~\cite{Traykov20084532}, operated at a frequency of 500 kHz with a peak-to-peak RF voltage of $V_{\textrm{RF}}=380$ V applied between neighboring rods; the opposite half-moon-shaped electrodes had a tip distance of 5 mm. For on-line optical spectroscopy, the ions were trapped at the end of the RFQ by suitable axial potentials (Paul trap); {\it cf.} Fig.~\ref{setup}. Typically $10^3$ $^{212}$Ra$^+$, $10^4$ $^{213}$Ra$^+$, and $10^2$ $^{214}$Ra$^+$ ions could be stored. The storage time was of order 100 seconds at a residual gas pressure of $10^{-8}$ mbar (the lifetimes for radioactive decay are 13 s, 164 s, and 2.5 s for $^{212}$Ra, $^{213}$Ra, and $^{214}$Ra, respectively). A N$_2$ or Ne buffer gas was used to aid effective catching and trapping of the radioactive particles from the beam in the RFQ. This gas dissipated the large (eV) energies of the ion beam, compressed the trapped cloud, and also enhanced the storage time. The buffer gas influenced the level lifetimes of the ions because of optical quenching and (hyper)fine-structure mixing of the metastable states. Based on Ref.~\cite{PhysRevA.58.264} it was expected that Ne had the smallest influence on the level lifetimes.

Home-built Extended Cavity Diode Lasers (ECDLs) were used to drive the optical transitions ({\it cf.} Fig.~ \ref{level}). Light to drive the \pump transition at wavelength $\lambda_1=468$ nm came from NDHA210APAE1 laser diodes from Nichia; the \ir transition at wavelength $\lambda_2=1079$ nm was driven with light from a LD-1080-0075-1 diode from Toptica; the \shelf line at wavelength $\lambda_3=708$ nm was excited with light from a HL7001MG diode from Opnext. The laser light was delivered to the ion trap with single-mode optical fibers. The beams were overlapped with polarizing beam splitters and a dichroic mirror and sent axially through the trap to minimize scattered light. They were focussed to 1 mm diameter at the trap location. Typical laser beam powers $P$ at the trap center were $P(\lambda_1)=300$ $\mu$W, $P(\lambda_2)=600$ $\mu$W, and $P(\lambda_3)=150$ $\mu$W.
The wavelengths were monitored with two High-Finesse Angstrom WS6 VIS and IR wavelength meters. Absolute frequency calibration for light at $\lambda_1$ was provided by an absorption line in Te$_2$ at wavelength 468.3185 nm (no. 178 in Ref.~\cite{telluriumatlas}) through linear absorption in a Te$_2$ glass cell at 450 K. Light at $\lambda_3$ was calibrated by linear absorption at the P(146)(2-8) resonance in I$_2$ in a cell at 500 K. Since for wavelength $\lambda_2$ no similar reference was available, it was determined with the IR wavelength meter. The IR wavelength meter was continuously cross-referenced with a high-finesse cavity. The transitions in Ra$^{+}$ were detected through fluorescence light from the \pump transition at wavelength $\lambda_1$. Because of the 10\% branching into the metastable \dt state, this fluorescence was only observed when both the \pump and \ir transitions were resonantly excited. The fluorescence light was imaged with a single lens of focal length $f = 30$ mm inside the vacuum through a low-pass filter with 80\% transmission for wavelengths shorter than 500 nm (Thorlabs FES0500) onto the photocathode of a photomultiplier (Hamamatsu R7449). The collection solid angle was 0.4 sr.

\begin{figure}[t]
\includegraphics[width = 5.0cm, angle = 90]{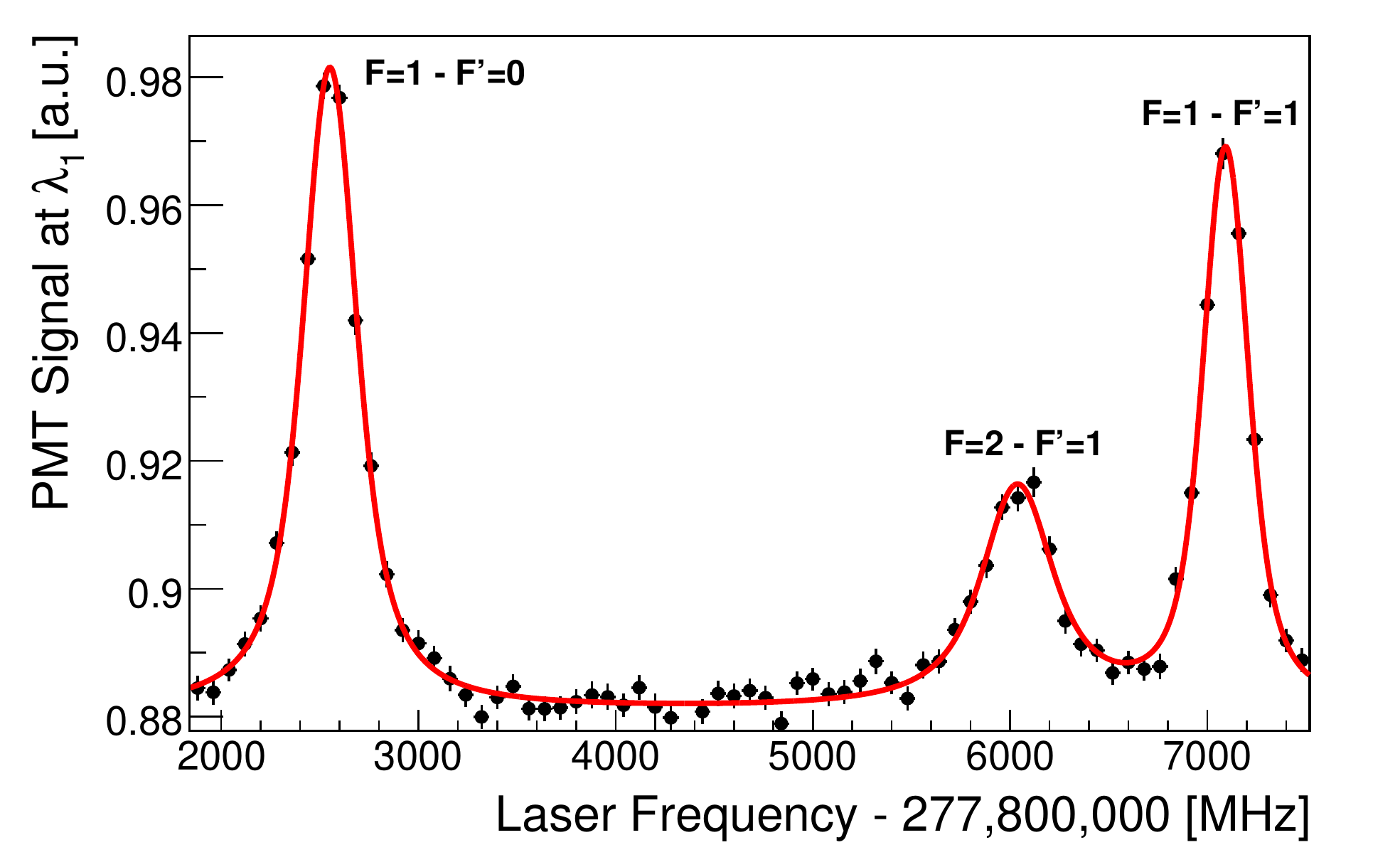}
\caption{HFS of the \ir transition in $^{213}$Ra$^+$. The solid line represents a fit of three Voigt profiles to the data. One parameter is used to fit the Gaussian widths of the three resonances, which yields a FWHM of 181(20) MHz. The Lorentzian widths (FWHM) are 245(20), 368(70), and 147(9) MHz (left to right). The different Lorentzian widths are due to saturation effects. The fit has a reduced $\chi^2 =1.1$ at 62 degrees of freedom. }\label{HFS}
\end{figure}

\begin{figure}[t]
\includegraphics[width = 5.0cm, angle = 90]{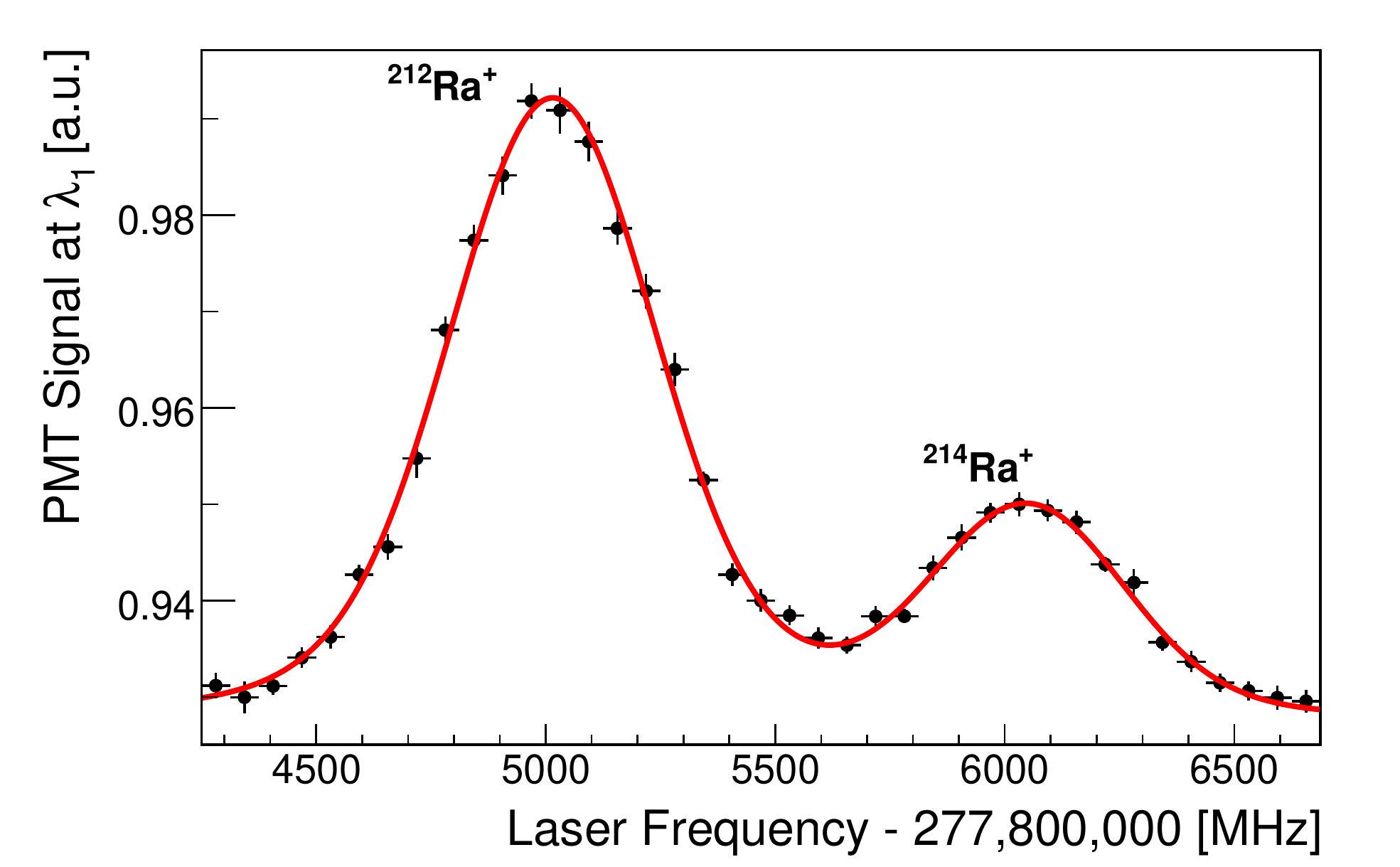}
\caption{\ir resonances in $^{212}$Ra$^+$ and $^{214}$Ra$^+$. The solid line represents a fit of two Voigt profiles to the data. One parameter is used to fit the Gaussian widths of the two resonances, which yields a FWHM  of 436(40) MHz. The Lorentzian widths (FWHM) are 201(50) MHz and 145(60) MHz (left to right). The fit has a reduced $\chi^2 = 0.92$ at $22$ degrees of freedom.} \label{IS1080}
\end{figure}

To study the HFS of the \ir transition in $^{213}$Ra$^+$ the wavelengths of the light from two diode lasers at $\lambda_1$ were kept close to the resonances \sh $F$=1 - \ph $F$'=0 and \sh $F$=0 - \ph $F'$=1. The frequency of the laser light at $\lambda_2$ was scanned over the resonances. For this measurement N$_2$ buffer gas was used. Collisions admixed the two hyperfine levels of the \dt level, ensuring that no significant shelving to the metastable \dt $F$=1 ($F$=2) state occurred when the \dt $F$=2 ($F$=1) was depopulated by the resonant laser light at $\lambda_2$. The frequency was calibrated with the IR wavelength meter. The measured line shapes are shown in Fig.~ \ref{HFS}. The different Lorentzian line-widths are due to saturation effects related to various relaxation rates \cite{AtomicPhysics}, here introduced by the buffer gas. The measured HFS splitting 4542(7) MHz for the \ph state is within 2 standard deviations of the value 4525(5) MHz obtained at ISOLDE~\cite{ISOLDE1987isohyper}. For the \dt state the HFS splitting is measured as 1055(10) MHz; the extracted \ph and \dt HFS constants $A$ are given in Table~\ref{tab:TableHFS}. The theoretical predictions \cite{wansbeekprar08, PhysRevA.79.062505} are in good agreement with the experimental values.
\begin{table}[t]
\caption{HFS constants $A$ [MHz] of the \ph and  \dt states in $^{213}$Ra$^+$. The most recent theoretical values were converted to $^{213}$Ra$^+$ by using the magnetic moment measured at ISOLDE~\cite{ISOLDEmomentswPhysRevLett.59.771}.}
\begin{ruledtabular}
\begin{tabular}{lll}
       & \ph & \dt \\ \hline
        This work & 4542(7) & 528(5) \\
        ISOLDE~\cite{ISOLDE1987isohyper} & 4525(5) & -- \\ \hline
        Theory~\cite{wansbeekprar08} & 4555\footnotemark[1] & 543\footnotemark[1] \\
        Theory~\cite{PhysRevA.79.062505} & 4565\footnotemark[1] & 541\footnotemark[1]  \footnotetext[1]{The theoretical uncertainty is at the \%-level \cite{timmermansError, safronovaError}.}\\
      \end{tabular}
\end{ruledtabular}
\label{tab:TableHFS}
\end{table}
The IS for the \ir transition of Ra$^+$ was obtained with light from two lasers kept close to wavelength $\lambda_1$. One of these laser beams excited the \pump \ transition in $^{212}$Ra$^+$, while the other one accessed either the \pump transition in $^{214}$Ra$^+$ or the \sh $F$=1 - \ph $F'$=0 transition in $^{213}$Ra$^+$. The frequency of the laser light at $\lambda_2$ was scanned over the \ir resonances of the isotopes under investigation. A typical spectrum of the \ir resonances of $^{212}$Ra$^+$ and $^{214}$Ra$^+$ is shown in Fig.~\ref{IS1080}. The IR wavelength meter was used for frequency calibration. In order to minimize the influence of the buffer gas on the resonance line shape, only Ne was used. The measurements were performed at gas pressures $3 \times 10^{-4}$, $3 \times 10^{-3}$, and $2 \times 10^{-2}$ mbar to study the influence of the buffer gas on the resonance line shapes. No significant effects on the measured IS were found. The resulting IS are summarized in Table \ref{tab:TableIS}.

\begin{figure}[t]
\includegraphics[width = 5.0cm, angle = 90]{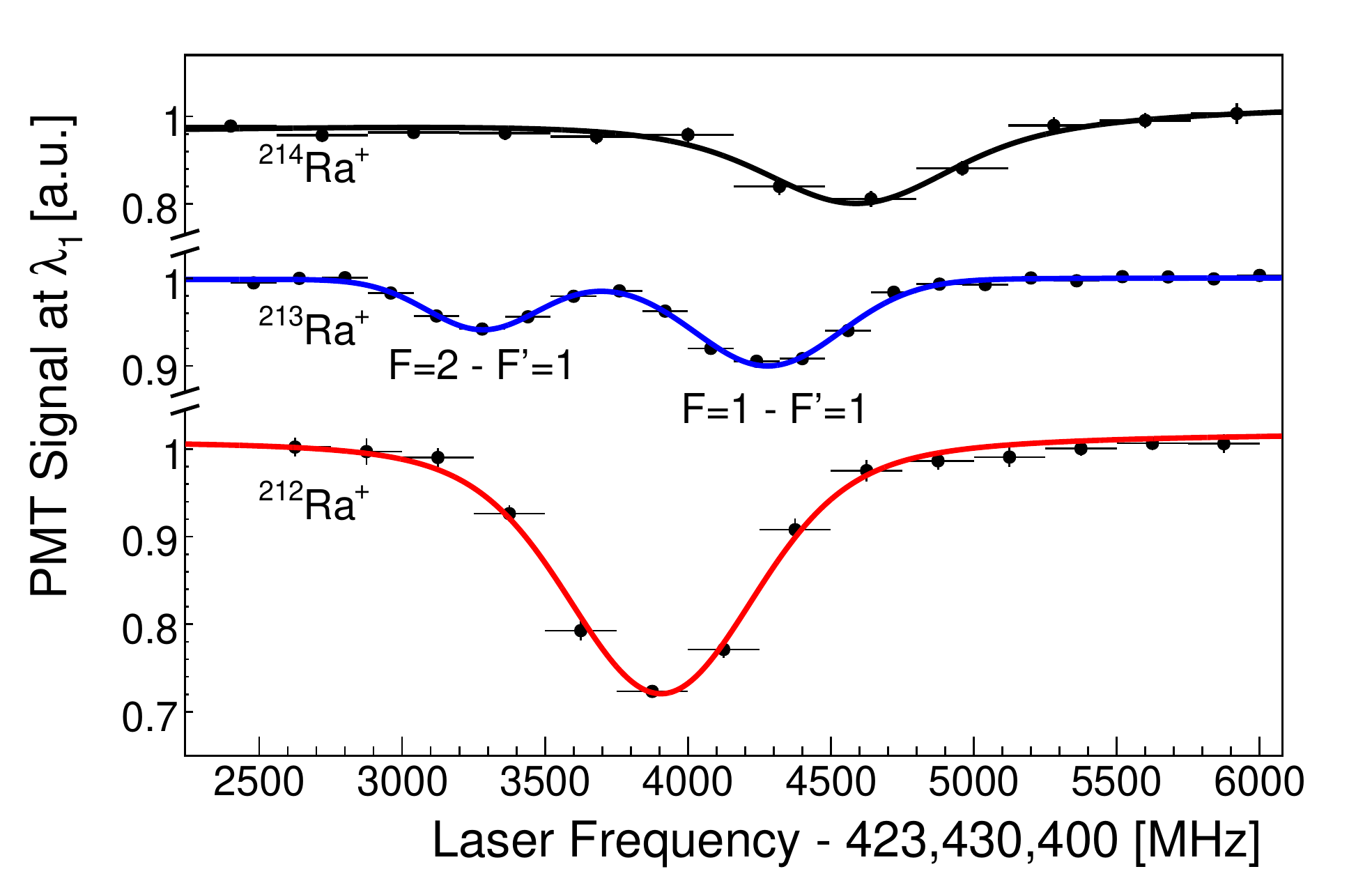}
\caption{Scans of the \shelf transitions. The solid lines represent fits of Voigt profiles to the data. The fits have reduced $\chi^2$'s of 0.81, 0.81, and 0.88 for $^{212}$Ra$^{+}$, $^{213}$Ra$^{+}$, and $^{214}$Ra$^{+}$, respectively, at 18, 26, and 17 degrees of freedom. The corresponding Gaussian and Lorentzian widths (FWHM) are 655(12) MHz and 243(10) MHz, respectively, for $^{212}$Ra$^{+}$, 363(30) and 144(23) MHz for the $F$=2 - $F'$=1 transition in $^{213}$Ra$^{+}$, and 451(270) MHz and 581(300) MHz for $^{214}$Ra$^{+}$. The different -temperature related- Gaussian widths are due to different neon buffer gas pressures. The different Lorentzian widths are caused by saturation effects which vary with gas pressure and laser power.} \label{IS708}
\end{figure}

\begin{figure}[t!]
\includegraphics[width = 5.0cm, angle = 90]{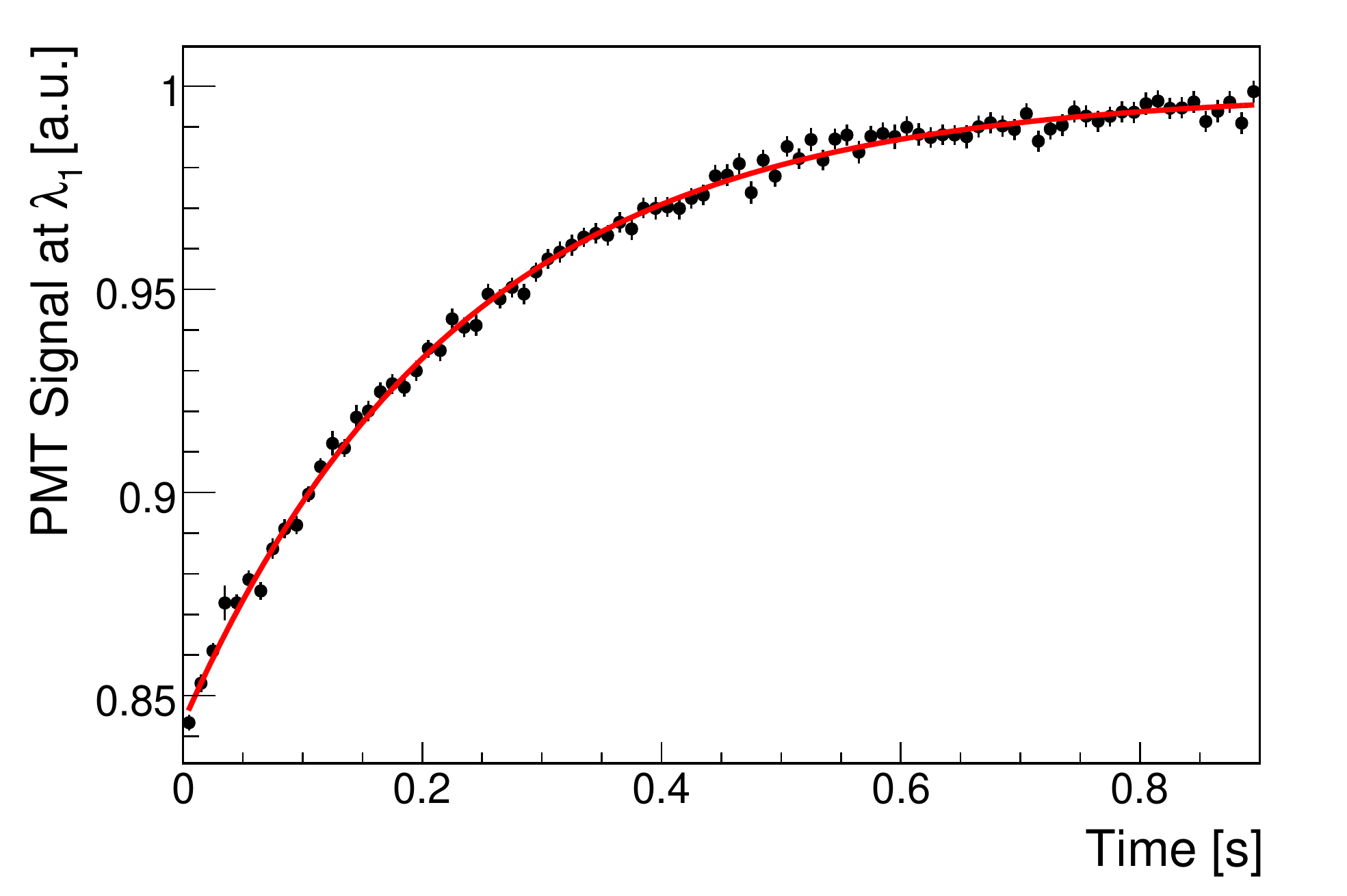}
\caption{Lifetime measurements of the $^{212}$Ra$^+$ \df state at a neon buffer gas pressure of $4 \times 10^{-5}$ mbar. The solid line represents a fit of an exponential function to the data. The fit yields a lifetime of 232(4) ms with a $\chi^2 =0.83$ at 87 degrees of freedom.} \label{lifetime}
\end{figure}

To determine the IS of the \shelf transition the lasers operating at $\lambda_1$ and $\lambda_2$ were kept close to resonance of a particular Ra$^+$ isotope. This created a fluorescence cycle. The frequency of the laser light at $\lambda_3$ was scanned over the resonances. Near resonance the ions were pumped to the \pt state, from which some 10\% decayed to the \df state ({\it cf.} Fig.~\ref{level}). In this metastable state the ions were shelved and did not participate in the fluorescence cycle. This caused a dip in the fluorescence signal, the position of which was calibrated against the P(146)(2-8) single-pass absorption resonance in molecular I$_{2}$ at $\nu_{\,\textrm{Iodine}}$=$423,433,720$ MHz. The scan linearity was verified with a high-finesse cavity. We found for the \shelf transition $\nu_{212}$=$\nu_{\,\textrm{Iodine}}+568(42)$ MHz for $^{212}$Ra$^+$ and $\nu_{214}$=$\nu_{\,\textrm{Iodine}}+1269(23)$ MHz for $^{214}$Ra$^{+}$.
\begin{table}
\caption{IS [MHz] of the \ir and \shelf transitions in Ra$^+$ isotope pairs. The measured HFS for the \dt state and a value extracted from Refs.~\cite{ISOLDE1987isohyper,ISOLDEmomentswPhysRevLett.59.771} for the \pt HFS were used to obtain the IS with respect to the center-of-mass of the resonances in $^{213}$Ra$^+$.}
\begin{ruledtabular}
\begin{tabular}{cccc}
    & $^{214}$Ra\,--\,$^{212}$Ra &  $^{213}$Ra\,--\,$^{212}$Ra & $^{214}$Ra\,--\,$^{213}$Ra  \\ \hline
\dt - \ph&  1032(5)  & 318(11) & 714(12) \\
\dt - \pt&  701(50)  & 248(50) & 453(34)
\end{tabular}
\end{ruledtabular}
\label{tab:TableIS}
\end{table}
For $^{213}$Ra$^+$ the fluorescence cycle was established by pumping on the \sh $F$=1 - \ph $F'$=0 transition and repumping on the \dt $F$=1 - \ph $F'$=0 transition. This left the \dt $F$=2 state largely depopulated. The frequency of the laser light at $\lambda_3$ was scanned over the resonances ({\it cf.} Fig.~\ref{IS708}). A rate equation model of the system that explicitly takes the (hyper)fine structure mixing caused by the Ne gas into account shows that a small mixing rate already causes several resonances to appear. The \dt $F$=1 - \pt $F'$=1 resonance is deformed by the close-lying \dt $F$=2 - \pt $F'$=2 transition. We use the \dt $F$=2 - \pt $F'$=1 resonance to determine the IS. The measurements were carried out at gas pressures $3 \times 10^{-4}$, $2 \times 10^{-3}$, and $2 \times 10^{-2}$ mbar. The power of the laser beam at $\lambda_3$ was varied between 50 and 150 $\mu$W; no significant changes were found. We found $\nu_{213}$=$\nu_{\textrm{Iodine}}-64(13)$ MHz. The measured isotope shifts are summarized in Table~\ref{tab:TableIS}. Ref.~\cite{rasmussenzp33} measured the absolute frequency of the \shelf transition in $^{226}$Ra$^{+}$ as 423,437,660(570) MHz. The value measured here for $^{212}$Ra$^{+}$is 423,434,288(42) MHz. This indicates an IS of 3.4(6) GHz between the two isotopes.
The demonstrated shelving to the \df state by accessing the \shelf transition also enables a measurement of the lifetime of this metastable state. The lasers at $\lambda_1$ and $\lambda_2$ were kept close to resonance in $^{212}$Ra$^+$, while the laser light at $\lambda_3$ was pulsed with 170 ms on-periods and 670 ms off-periods by a mechanical chopper wheel. The laser light at $\lambda_3$ was kept on resonance to populate \df via the \pt state. When the laser light at $\lambda_3$ was switched off, the \df state depopulated and the ions re-entered the fluorescence cycle with a time constant equal to the lifetime of the \df state (\emph{cf.} Fig.~\ref{lifetime}).
However, the neon buffer gas caused a reduction of the lifetime of the metastable state by quenching it to the ground state. To estimate the effects of the buffer gas, measurements were conducted at different gas pressures ranging from $10^{-2}$ to $10^{-5}$ mbar. The buffer gas was shown to have a strong influence on the optical lifetime. However, no detailed theory for this system is presently available to extrapolate the lifetime to zero pressure. A lower bound on the radiative lifetime of the \df state was found to be 232(4) ms; it corresponds to the lifetime measured at the lowest pressure of some $4 \times 10^{-5}$ mbar. Corrections for the radioactive lifetime of $^{212}$Ra and for the replacement time can be neglected. Theoretical predictions are 297(4) ms~\cite{sahoopra07} and 303(4) ms~\cite{PhysRevA.79.062505}. Our experimental result is an important confirmation that the \df state is indeed long-lived. This is a necessary property in view of the long coherence times needed in APV experiments with a single trapped ion~\cite{fortsonprl93}.\\
In conclusion, for the first time on-line excited-state laser spectroscopy was performed on short-lived trapped ions. HFS measurements are suited to test wave functions at the origin, whereas the measurement of radiative lifetimes test them at larger distances. IS measurements probe atomic theory and yield information about the size and shape of the atomic nucleus. These measurements test the atomic theory, the accuracy of which is indispensable for upcoming single-ion APV experiments aiming at an improved low-energy determination of the electroweak mixing angle~\cite{wansbeekprar08}. For the refinement of this test, Ra offers a chain of isotopes, where no measurements have been made and where theory is challenged to provide unbiased predictions.\\
We acknowledge the support received from the AGOR cyclotron group and the KVI technical personnel. O. B\"oll, O. Dermois, and L. Huisman were essential in the design and the setup of the experiment. We thank R. Hoekstra for useful discussions. This research was supported by the Stichting voor Fundamenteel Onderzoek der Materie (FOM) under Program 114 (TRI$\mu$P) and FOM projectruimte 06PR2499. O.O.V. acknowledges funding from the NWO Toptalent program.
\bibliography{thebib}
\end{document}